\documentclass[conference]{IEEEtran}
\IEEEoverridecommandlockouts
\usepackage{cite}
\usepackage{amsmath,amssymb,amsfonts}
\usepackage{algorithmic}
\usepackage{graphicx}
\usepackage{textcomp}
\usepackage{xcolor}
\def\BibTeX{{\rm B\kern-.05em{\sc i\kern-.025em b}\kern-.08em
    T\kern-.1667em\lower.7ex\hbox{E}\kern-.125emX}}

\usepackage{amsthm}
\newcommand{\RNum}[1]{\uppercase\expandafter{\romannumeral #1\relax}}
\usepackage{subfigure}

\newtheorem{theorem}{Theorem}[section]
    
\newtheorem{lemma}[theorem]{Lemma}
\theoremstyle{remark}

\begin{document}

\title{Reputation and Audit Bit Based Distributed Detection in the Presence of Byzantines}

\author{\IEEEauthorblockN{Chen Quan$^{1}$, Yunghsiang S. Han$^{2}$, Baocheng Geng$^{3}$, Pramod K. Varshney$^{1}$}

\IEEEauthorblockA{
\textit{$^{1}$Syracuse University\ $^{2}$University of Electronic Science and Technology of China \ $^{3}$University of Alabama at Birmingham }
}
}

\maketitle

\IEEEpeerreviewmaketitle
\begin{abstract}
    	In this paper, two reputation based algorithms called Reputation and audit based clustering (RAC) algorithm and Reputation and audit based clustering with auxiliary anchor node (RACA) algorithm are proposed to defend against Byzantine attacks in distributed detection networks when the fusion center (FC) has no prior knowledge of the attacking strategy of Byzantine nodes. By updating the reputation index of the sensors in cluster-based networks, the system can accurately identify Byzantine nodes. The simulation results show that both proposed algorithms have superior detection performance compared with other algorithms. The proposed RACA algorithm works well even when the number of Byzantine nodes exceeds half of the total number of sensors in the network. Furthermore, the robustness of our proposed algorithms is evaluated in a dynamically changing scenario, where the attacking parameters change over time.  We show that our algorithms can still achieve superior detection performance.
\end{abstract}
\section{Introduction}
In distributed wireless sensor networks (WSNs), local sensors send their decisions regarding the presence or absence of the phenomenon of interest (PoI) to the fusion center (FC) and the FC makes a final decision regarding the presence or absence of the PoI. 
 Due to its energy-efficiency, distributed framework is widely adopted in many bandwidth-limited scenarios, e.g., IoT, cognitive radio networks and military surveillance systems. However, the open nature of  WSNs makes the distributed system vulnerable to various attacks such as Byzantine attacks\cite{zhang2015byzantine,lamport2019byzantine}, wiretap, jamming and spoofing \cite{jover2014enhancing,gai2017spoofing}. In this paper, we focus on Byzantine attacks where the sensors in a network may be compromised and controlled by adversaries and send falsified decisions to the FC.

The Byzantine attack problem in distributed detection systems has been studied in the literature \cite{sonnek2007adaptive,zeng2010reputation,rawat2010collaborative,kailkhura2013covert,vempaty2014false,kailkhura2015distributed,hashlamoun2017mitigation,hashlamoun2018audit,quan2022enhanced}. For example, in order to enhance detection performance and improve robustness, some reputation based schemes have been proposed in spectrum sensing networks \cite{sonnek2007adaptive,zeng2010reputation}. In \cite{kailkhura2015distributed}, the optimal attacking strategies are analyzed for two cases where the FC has knowledge of the attackers' strategy and where the FC does not know the attackers' strategy.
Moreover, the authors in \cite{hashlamoun2017mitigation,hashlamoun2018audit,quan2022enhanced} developed the audit bit based distributed detection scheme to combat Byzantines, where each sensor sends one additional audit bit to the FC which gives more information about the behavioral identity of each sensor.



Different from previous works, we consider that the FC does not have prior knowledge of the attacking strategy of Byzantine nodes, namely the flipping probabilities. We propose two reputation based algorithms to mitigate the effect of Byzantine attacks. In both proposed algorithms, we utilize the reputation indexes of sensors to represent the trustworthiness of sensors in the network. The reputation indexes of sensors are updated at each time step according to their behaviors. Sensors with low reputation indexes are usually identified as Byzantine nodes and are excluded from the decision-making process. In particular, the audit bit based mechanism and the Partitioning Around Medoid (PAM) algorithm are developed to update the reputation indexes of sensors in the network and to identify potential Byzantine nodes. 
The ability to identify Byzantine nodes can be further enhanced by the use of anchor nodes even when the number of Byzantine nodes exceeds half of the total number of sensors in the networks. The robustness of proposed algorithms is tested both in dynamic (attacking parameters change dynamically over time) and static (attacking parameters remain the same) scenarios. Simulation results show that our proposed algorithms are capable of defending against attackers
in both scenarios.


\section{System model}

Consider a binary hypothesis testing problem with the two hypotheses denoted by ${H}_0$ and ${H}_1$.
A WSN is comprised of $N$ sensors and one FC, where the FC makes a final decision on which  hypothesis is true based on the sensor's local decisions. The sensors are divided into groups of two and there are a total of $G=N/2$ groups in the network. The sensors make binary decisions on whether $H_0$ or $H_1$ is true by utilizing the likelihood ratio (LR) test. For ease of notation, let us assume that each sensor has the same probabilities of detection and false alarm, i.e., $P_d=P(v_i=1|\mathcal{H}_1)$ and $P_f=P(v_i=1|\mathcal{H}_0)$ for $i\in\{1,\ldots, N\}$, where $v_i$ is the decision made by sensor $i$. In addition to its local decision $v_i$, each sensor also sends one more decision, which comes from its group member, to the FC and we call this additional decision the audit bit as described in \cite{hashlamoun2017mitigation,hashlamoun2018audit,quan2022enhanced}.

\begin{figure*}[htb]
  \centering
  \subfigure[The architecture of group k.]{
    \includegraphics[width=10em,height=10em]{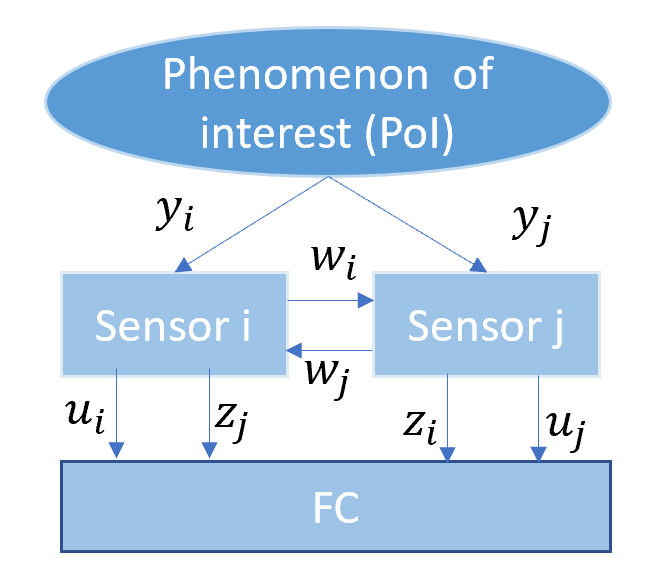}}
      \subfigure[The block diagram of the proposed algorithm.]{
    \includegraphics[width=32em,height=6em]{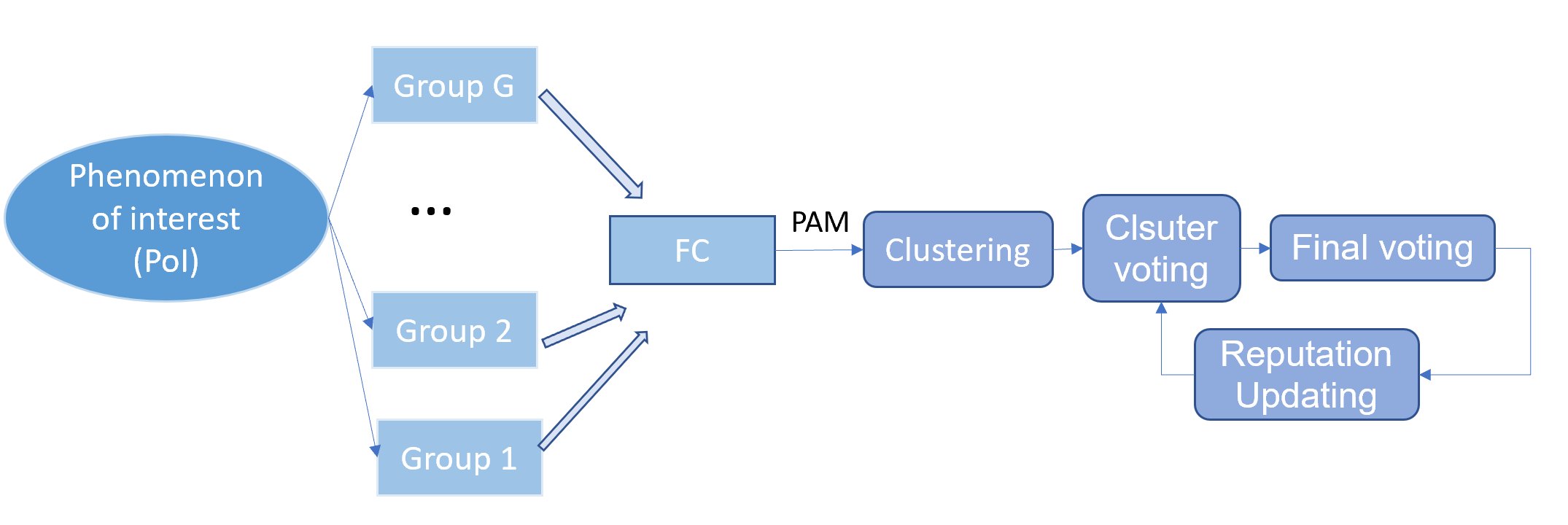}}
  \caption{The architecture of any group k is shown in (a). The block diagram of the proposed algorithm is shown in (b).}
  \label{Fig.main} 
\end{figure*}

For simplicity, let $i$ and $j$ represent the sensors in the same group. As shown in Fig.~\ref{Fig.main}(a), after making its own decision $v_i$, sensor $i$ sends (i) $u_{i}$ directly to the FC; (ii) $w_{i}$ to the sensor $j$ in the same group; (iii) $z_{j}$, corresponding to $w_j$ coming from the sensor $j$ in the same group, to the FC. Similarly, sensor $j$ also sends two decisions $u_{j}$ and $z_{i}$ to the FC.
If sensor $i$ is a Byzantine node, i.e., $i= B$, the decisions $v_i$, $ w_i$ and $u_i$ are not necessarily the same and $z_j$ are also not necessarily equal to $u_j$. If sensor $i$ is honest, i.e., $i= H$, it sends genuine or uncorrupted information to the FC. Hence, given a Byzantine node $i$, the attacking parameters $p_1$ and $p_2$ are given by $p_1=p(v_i\neq u_i|i=B)=p(v_i\neq w_i|i=B)$ and $p_2=p(w_j\neq z_j|i=B)$.
Given an honest node $i$, we have $p_1=p(v_i\neq u_i|i=B)=p(v_i\neq w_i|i=B)=0$ and $p_2=p(w_j\neq z_j|i=B)=0$.
In other words, $p_1$ and $p_2$ represent the probabilities that a node flips its own decision and flips the decision coming from its group member, respectively. If $u_i=z_i$, we have a `match' for sensor $j$, otherwise, we have a `mismatch' for sensor $j$. Similarly, for sensor $i$, we have a `match' when $u_j=z_j$ and a `mismatch' when $u_j\neq z_j$. We assume that a fraction $\alpha_0$ of the $N$ sensors are Byzantine nodes and they attack independently. The FC is not aware of the identity or the attacking strategy of Byzantine nodes in the network. Hence, each node has the probability $\alpha_0$ to be a Byzantine node and the FC does not know the values of $p_1$ and $p_2$. We consider the more general and practical attacking strategy as stated in \cite{quan2022enhanced} which allows the Byzantines to be strategic by optimally employing unequal probabilities $p_1$ and $p_2$.

\section{Proposed Reputation and Audit Bit Based Clustering Algorithms}
In this section, we present the proposed robust defense algorithms for the system under attack when the FC does not possess the knowledge of the attacking strategy, namely $p_1$ and $p_2$, used by Byzantine nodes. We also evaluate the performance of our proposed algorithms in this section. 
\subsection{Proposed Algorithms}
Upon receiving measurements $\{u_i\}_{i=1}^N$ and $\{z_i\}_{i=1}^N$, the FC is able to determine the match and mismatch (MMS) results ($u_i=z_i$ or $u_i\neq z_i$) for all the sensors in the network. Based on the received measurements and the MMS results, we propose a robust reputation based algorithm to defend against Byzantine attacks. The proposed reputation and audit bit based clustering (RAC) algorithm consists of four successive phases and the flow chart is shown in Fig.\ref{Fig.main}(b).

\paragraph{Macro clustering phase} At time step $t$, $T$ most recent decisions of each node are utilized. The FC keeps a $N(2T+1)$ dimensional vector\footnote{The information of each sensor consists of $T$ local decisions, $T$ MMS results, and one reputation index.} to store the information corresponding to each sensor in the network, which consists of the records of local decisions, the records of MMS results and the updated reputation index. 
We make use of the MMS results to cluster or partition the sensors into two different sets, which are $\underline{\mathcal{T}}$ and $\overline{\mathcal{T}}$. If the MMS results for both sensors in the same group are always 'match', the sensors in this group are placed in set $\underline{\mathcal{T}}$, otherwise, they are placed in set $\overline{\mathcal{T}}$.
\paragraph{Micro clustering phase}
After partitioning all the sensors into two sets, in each set, we employ the Partitioning Around Medoid (PAM) algorithm\footnote{PAM is one possible algorithm to implement K-medoid clustering. K-medoid clustering is a prominent clustering technique which attempts to minimize the distance between points assigned to a cluster and a point designated as the center of that cluster, namely the medoid of that cluster.}\cite{kaufman2009finding} to partition the sensors in the same set into several clusters or subsets based on the decisions $\{\mathbf{u}_i\}_{i=1}^N$. We assume that the sensors are grouped into $K$ clusters in each set via PAM.\footnote{The number of clusters in different sets are assumed to be the same for simplicity.} Hence, we have a total $2K$ clusters in the network.
\paragraph{Voting phase}
The Voting phase contains two successive steps, i.e., Intra-cluster voting and Inter-cluster voting.
\paragraph*{\underline{Intra-cluster voting}}
After each sensor makes the decision at time step $t$, we perform cluster voting by weighting the decisions of the sensors in that cluster with their impact factors. 
Let $I_i(t)$ denote the impact factor of sensor $i$ at time step $t$ that is given as
\begin{align}
    I_i(t)=\left\{
    \begin{array}{rcl}
        \frac{1}{d_{t}(i,m_k)}&\text{if sensor $i$ is in cluster $k$}\\
        0& \text{if sensor $i$ is not in cluster $k$}
    \end{array}\right.
\end{align}
where $d_{t}(i,m_k)$ denotes the Hamming distance between the decision record of sensor $i$ and the decision record of the medoid of that cluster for the most recent $T$ time steps, and, $m_k$ represents the index of the medoid of cluster $k\in\{1,\dots,2K\}$. Note that the first $K$ clusters consist of the sensors in set $\underline{\mathcal{T}}$ and the rest of the clusters consist of the sensors in set $\overline{\mathcal{T}}$. The cluster vote for cluster $k$ at time $t$ is given by
\begin{equation}\label{eq:cluster_vote}
    \mathcal{V}_k(t)=\left\lfloor \frac{\sum_{i=1}^{N}I_i(t)u_i(t)}{\sum_{i=1}^{N}I_i(t)}\right\rceil,
\end{equation}
where $\lfloor x \rceil$ means rounding $x$ to the nearest integer. $\mathcal{V}_k(t)$ is either 0 or 1 to represent the absence or presence of the PoI, respectively, at the cluster level.
\paragraph*{\underline{Inter-cluster voting}}
After receiving all the cluster decisions, the FC makes a decision regarding the behavioral status, i. e., Byzantine or not, of each cluster based on the cluster reputation index.  The cluster reputation index is defined as the averaged reputation indexes of the sensors in that cluster. Initially, we assume that the sensors are equally trustworthy. We, therefore, set the initial reputation indexes for all sensors in the network to be $r_{init}$. Based on reputation indexes of all the sensors, we are able to calculate the cluster reputation indexes for all clusters. If the cluster reputation index is below a threshold $\lambda_{valid}$, the cluster is temporarily considered to be Byzantine and the cluster decision from that cluster is not taken into consideration when the FC makes the final decision regarding the hypothesis that is true. The decision rule used by the FC is expressed as
\begin{equation}
    \gamma_1\sum_{k=1}^K\beta_k\mathcal{V}_k(t)+\gamma_2\sum_{k=K+1}^{2K}\beta_k\mathcal{V}_k(t)\overset{v_{fc}(t)=1}{\underset{v_{fc}(t)=0}{\gtrless}}\lambda_{fc},
\end{equation}
where $\lambda_{fc}$ is the threshold used by the FC, $v_{fc}(t)$ is the final decision at time step $t$, and, $\gamma_1$ and $\gamma_2$ are the weights of the cluster decisions for the first $K$ clusters (clusters in set $\underline{\mathcal{T}}$) and the ones for the rest of the clusters, respectively. Based on Lemma~\ref{lemma1}, we give appropriate values to $\gamma_1$ and $\gamma_2$ to emphasize different importance of the cluster decisions from different sets. $\beta_k$ is the weight of decisions from cluster $k$ in the corresponding set and it is given by
\begin{equation}
    \beta_k=\left\{
    \begin{array}{rcl}
       \frac{Y_k}{\sum_{k=1}^{K}Y_k}&\text{if cluster $k\in\{1,2,\dots,K\}$}\\
        \frac{Y_k}{\sum_{k=K+1}^{2K}Y_k}&\text{if cluster $k\in\{K+1,K+2,\dots,2K\}$}
    \end{array}\right.
\end{equation}
where $Y_k=\frac{n_kF_k}{\sum_{k=1}^{2K}n_kF_k}$, $n_k$ is the number of sensors in cluster $k$ and $F_k$ is the cluster behavioral identity indicator for cluster $k$. $F_k=1$ represents the fact that the cluster $k$ is not considered Byzantine and $F_k=0$ represents that the cluster $k$ is considered Byzantine. We set $\gamma_1>\gamma_2$ to emphasize that the importance of cluster decisions coming from set $\underline{\mathcal{T}}$ is greater than the ones coming from set $\overline{\mathcal{T}}$ according to Lemma~\ref{lemma1}.
\begin{lemma}\label{lemma1}
The sensors in set $\underline{\mathcal{T}}$ have higher probabilities to be Byzantine nodes than the sensors in set $\overline{\mathcal{T}}$ if $\alpha_0\leq0.8$.
\end{lemma}

\begin{IEEEproof}
The proof is relegated to Appendix \ref{proof_lemma}.
\end{IEEEproof}
\paragraph{Reputation updating phase}
At the end of each time step, the reputation indexes of all the sensors are updated. The final decision of the FC is propagated back to the clusters and further to the individual sensors to update the reputation index of each sensor. If the final decision is the same as a cluster decision, that cluster gets a positive feedback; otherwise, it gets a negative feedback. Similarly, if the cluster decision is the same as a sensor decision in that cluster, that sensor gets a positive feedback; otherwise, it gets a negative feedback. The reputation updating rule of sensor $i$ is given as
\begin{equation}\label{eq:repu_1}
    r_i=r_i+M(v_{fc}(t),\mathcal{V}_k(t))g_k\frac{H_i(t)}{\sum_{i=1}^NH_i(t)},
\end{equation}
where $g_k=\frac{\sum_{i=1}^NM(u_i(t),\mathcal{V}_k(t))I_i(t)}{\sum_{i=1}^NI_i(t)}$
represents the step size to penalize or reward a sensor in cluster $k$ and $M(a,b)$ is an indicator function that returns 1 if $a$ equals $b$ and returns -1 otherwise. Let $H_i(t)$ denote the reputation impact factor of sensor $i$ at time step $t$ and it is given by
\begin{equation}
    H_i(t)=\left\{
    \begin{array}{rcl}
        \frac{1}{D_{t}(i,m_k)}&\text{if sensor $i$ is in cluster $k$}\\
        0&\text{if sensor $i$ is not in cluster $k$}
    \end{array}\right.
\end{equation}
where $D_{t}(i,m_k)$ is the Hamming distance between the MMS result record of sensor $i$ and the MMS result record of the medoid of that cluster for the most recent $T$ time steps. The reputation index of cluster $k$ is updated via $R_k=\frac{\sum_{i\in\mathcal{E}_k}r_i}{n_k}$
for $k=1,2,\dots,K$, where $\mathcal{E}_k$ is the set of indices of the sensors in cluster $k$. If $R_k$ is smaller than a threshold $\tau$, we temporarily remove all the sensors in cluster $k$ and repeat step c).\footnote{Because it is possible that several honest nodes are grouped into a Byzantine cluster or the cluster is wrongly identified as Byzantine, we just temporarily remove all the sensors in that cluster.}

However, as we will see later, the simulation results in Fig. 3 shows that the system employing the RAC algorithm breaks down when the Byzantine nodes adopt the strategy that $p_1$ approaches 1, $p_2$  approaches 0 and $\alpha_0\geq0.5$. Hence, we further propose an algorithm with auxiliary anchor nodes to overcome that problem. We use the same procedure as in the above algorithm except the Reputation updating phase. Assume there are $J$ ($J\ll N$) anchor nodes in the network which can be trusted by the FC, and $P_d$ and $P_f$ are the same as the other sensors in the network. Let $A(t)$ represent the final decision according to the local decisions from anchor nodes at time step $t$ and the final decision is decided by a majority vote if more than 1 anchor node is used. The reputation updating rule of sensor $i$ given in \eqref{eq:repu_1} is reformulated as $r_i=r_i+M(v_{fc}(t),\mathcal{V}_k(t))g_kf\frac{H_i(t)}{\sum_{i=1}^NH_i(t)}$,
with auxiliary anchor node, where $f=\frac{\sum_{q=t-T+1}^{t}Q(A(t),A(q))}{T}M(A(t),v_{fc}(t))$. $f$ can be regarded as a reward (or punishment) step size of the reputation index when the decision of the anchor node is the same as the final decision (or different from the final decision). $Q(a,b)$ is an indicator function that returns 1 if $a$ equals $b$ and returns 0 otherwise. 

\subsection{Performance Analysis}
We evaluate the attacking strategy of Byzantines that makes the FC totally blind when we employ our proposed algorithms. Since Macro clustering is performed, we need to consider two cases: (i) The sensors are in set $\underline{\mathcal{T}}$; (ii) The sensors are in set $\overline{\mathcal{T}}$. The probabilities of detection and false alarm are different for the sensors in different sets. Let $\underline{\pi}_{11}$, $\underline{\pi}_{10}$, ans $\overline{\pi}_{11}$, $\overline{\pi}_{10}$ denote the probabilities of detection and false alarm for the sensors in set $\underline{\mathcal{T}}$ and the sensors in set $\overline{\mathcal{T}}$, respectively. We have $\underline{\pi}_{11}=P_d(1-\underline{\alpha}p_1)+\underline{\alpha}p_1(1-P_d)$, $\underline{\pi}_{10}=P_f(1-\underline{\alpha}p_1)+\underline{\alpha}p_1(1-P_f)$, $\overline{\pi}_{11}=P_d(1-\overline{\alpha}p_1)+\overline{\alpha}p_1(1-P_d)$ and $\overline{\pi}_{10}=P_f(1-\overline{\alpha}p_1)+\overline{\alpha}p_1(1-P_f)$. $\underline{\alpha}$ is the probability that one sensor in set $\underline{\mathcal{T}}$ is a Byzantine node and $\overline{\alpha}$ is the probability that one sensor in set $\overline{\mathcal{T}}$ is a Byzantine node and they are given by
\begin{subequations}\label{eq:alpha_LL}
\begin{align}
    \underline{\alpha}=&\frac{\alpha_0^2f_{1}+\alpha_0(1-\alpha_0)f_{2}}{\alpha_0^2f_{1}+2\alpha_0(1-\alpha_0)f_{2}+(1-\alpha_0)^2},\\
    \overline{\alpha}=&\frac{\alpha_0-(\alpha_0^2f_{1}+\alpha_0(1-\alpha_0)f_{2})}{1-(\alpha_0^2f_{1}+2\alpha_0(1-\alpha_0)f_{2}+(1-\alpha_0)^2)},
\end{align}
\end{subequations}
where $f_{1}=[2p_1p_2(1-p_1)+(1-2p_1+2p_1^2)(1-p_2)]^2$ and $f_{2}=f_{BH}^{(1)}=(1-p_2)(1-2p_1+2p_1^2)$. To totally blind the FC, the adversaries need to ensure that the following equalities simultaneously hold.
\begin{subequations}\label{eq:kld2}
    \begin{align}
        D\left(\underline{\alpha},p_1,p_2\right)&=0\label{eq:kld2_a}\\
        D\left(\overline{\alpha},p_1,p_2\right)&=0\label{eq:kld2_b}
    \end{align}
\end{subequations}
where $D(\cdot)$ is the Kullback–Leibler divergence (KLD), and $D\left(\underline{\alpha},p_1,p_2\right)=\underline{\pi}_{11}\log\frac{\underline{\pi}_{11}}{\underline{\pi}_{10}}+\underline{\pi}_{01}\log\frac{\underline{\pi}_{01}}{\underline{\pi}_{00}}$ and $D\left(\overline{\alpha},p_1,p_2\right)=\overline{\pi}_{11}\log\frac{\overline{\pi}_{11}}{\overline{\pi}_{10}}+\overline{\pi}_{01}\log\frac{\overline{\pi}_{01}}{\overline{\pi}_{00}}$ . The equations \eqref{eq:kld2_a} and \eqref{eq:kld2_b} always hold only when $\underline{\pi}_{11}=\underline{\pi}_{10}$ and $\overline{\pi}_{11}=\overline{\pi}_{10}$, respectively, which yields $\underline{\alpha}p_1=\frac{1}{2}$ and $\overline{\alpha}p_1=\frac{1}{2}$. Moreover, since we assign different weights to different
cluster decisions, i.e., $\gamma_1$ and $\gamma_2$, according to the fusion rule shown in \eqref{fusion_rule}, the optimal attacking strategy is $p_1=1$, $p_2=0$, and $\alpha_0=0.5$.

Let $P_{HH}^{diff}$ and $P_{BH}^{diff}$ denote the probabilities that two honest nodes differ in their sensing reports and the probability that one honest node and one Byzantine node differ in their sensing reports, respectively, at any time step $t$. Obviously, $P_{HH}^{diff}$ and $P_{BH}^{diff}$ are given as
\begin{align*}
    &P_{HH}^{diff}=2\pi_0P_f(1-P_f)+2\pi_1P_d(1-P_d)\\
    &P_{BH}^{diff}=\pi_0[\kappa_{10}(1-P_f)+\kappa_{00}P_f]+\pi_1[\kappa_{11}(1-P_d)+\kappa_{01}P_d]
\end{align*}
where $\kappa_{10}=(1-P_f)p_1+P_f(1-p_1)$ and $\kappa_{11}=(1-P_d)p_1+P_d(1-p_1)$ are the probabilities of detection and false alarm for the Byzantine nodes in the network. If the adversaries want to totally deceive the FC so that the FC misplaces Byzantine nodes and honest nodes in the same cluster in the Micro clustering phase, we should have
\begin{equation}\label{eq:kld}
    D\left(P_{HH}^{diff}|P_{BH}^{diff}\right)=P_{HH}^{diff}\log_2\left({P_{HH}^{diff}}/{P_{BH}^{diff}}\right)=0
\end{equation}
where $D\left(P_{HH}^{diff}|P_{BH}^{diff}\right)$ is the KLD. The solutions of equation \eqref{eq:kld} are $p_1=0$ or $P_d=P_f=\frac{1}{2}$ which means that the adversaries can totally deceive the FC in the Micro clustering phase only when $p_1=0$ or $P_d=P_f=\frac{1}{2}$.


So in conclusion, the proposed mechanism pushes the Byzantine nodes to choose a large $p_1$ and a small $p_2$ to blind the FC in the Macro clustering phase. It is due to the fact that a small $p_2$ increases the probability that the Byzantine nodes are placed into set $\underline{\mathcal{T}}$, whose sensor decisions have more impact on the final decision. However, a large $p_1$ also increases the exposure to our defense mechanism in the Micro clustering phase which guarantees a good detection performance. Benefiting from our proposed scheme, we are also able to achieve a good detection performance even when $\alpha_0\geq 0.5$. It should also be noted that in prior work (for e.g.,\cite{rawat2010collaborative,vempaty2014false,hashlamoun2017mitigation,hashlamoun2018audit,kailkhura2014performance}), the FC can be made blind with only $50\%$ of Byzantine nodes in the network.

\section{Simulation Results and Discussion}
Some numerical results are presented in this section. We assume that identical sensors are utilized in the networks. Hence, we have $P_{d}=0.9$, $P_{f}=0.1$ for sensor $i\in\{1,\dots,N\}$ and anchor node $j\in\{1,\dots,J\}$. We set $N=500$, $r_{init}=0.5$, $\lambda_{valid}=0.5$, $\tau=0.5$, $\gamma_1=1.5$, $\gamma_2=0.5$ and $\lambda_{fc}=1$. 
The starting dimension $T$ is set to be 20 in this paper.

\begin{figure}[htb]
  \centering
    \includegraphics[width=17em,height=20em]{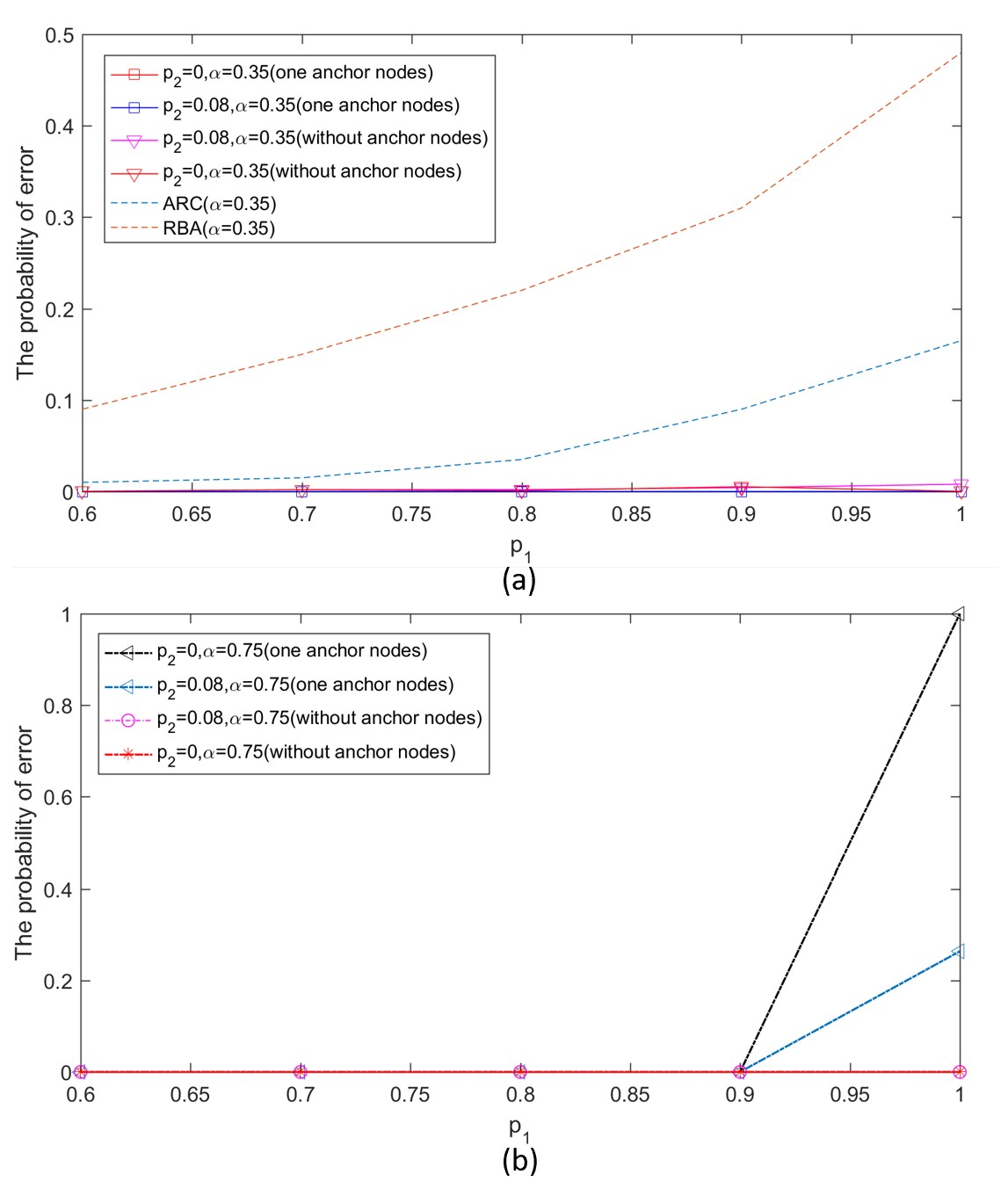}
  \caption{The probability of error versus $p_1$ given different value of $p_2$ and $\alpha_0$ for the proposed algorithms with one anchor node and without anchor nodes (i.e., RACA and RAC).}
  \label{Fig.main5}
\end{figure}

\begin{figure}[htb]
  \centering
    \includegraphics[width=17em,height=20em]{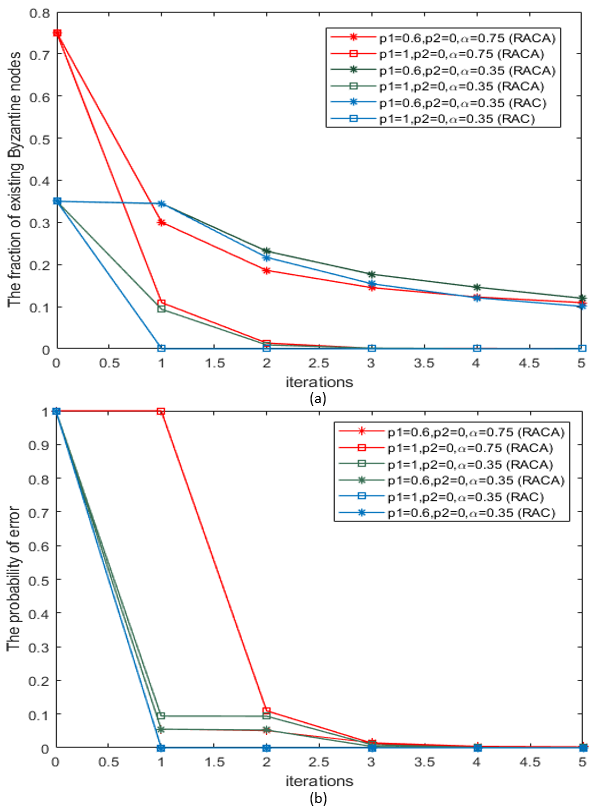}
  \caption{The probability of error and the fraction of identified Byzantine nodes versus $p_1$ for the proposed algorithms with one anchor node given $\alpha_0=0.35$ and $\alpha_0=0.75$.}
  \label{Fig.main1}
\end{figure}

\begin{figure}[htb]
  \centering{
    \includegraphics[width=17em,height=10em]{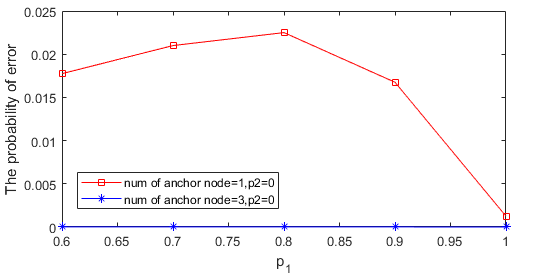}}
  \caption{The probability of error versus $p_1$ for the proposed RACA with different number of anchor nodes.}
  \label{Fig.main2} 
\end{figure}

\begin{figure}[htb]
  \centering
    \includegraphics[width=17em,height=10em]{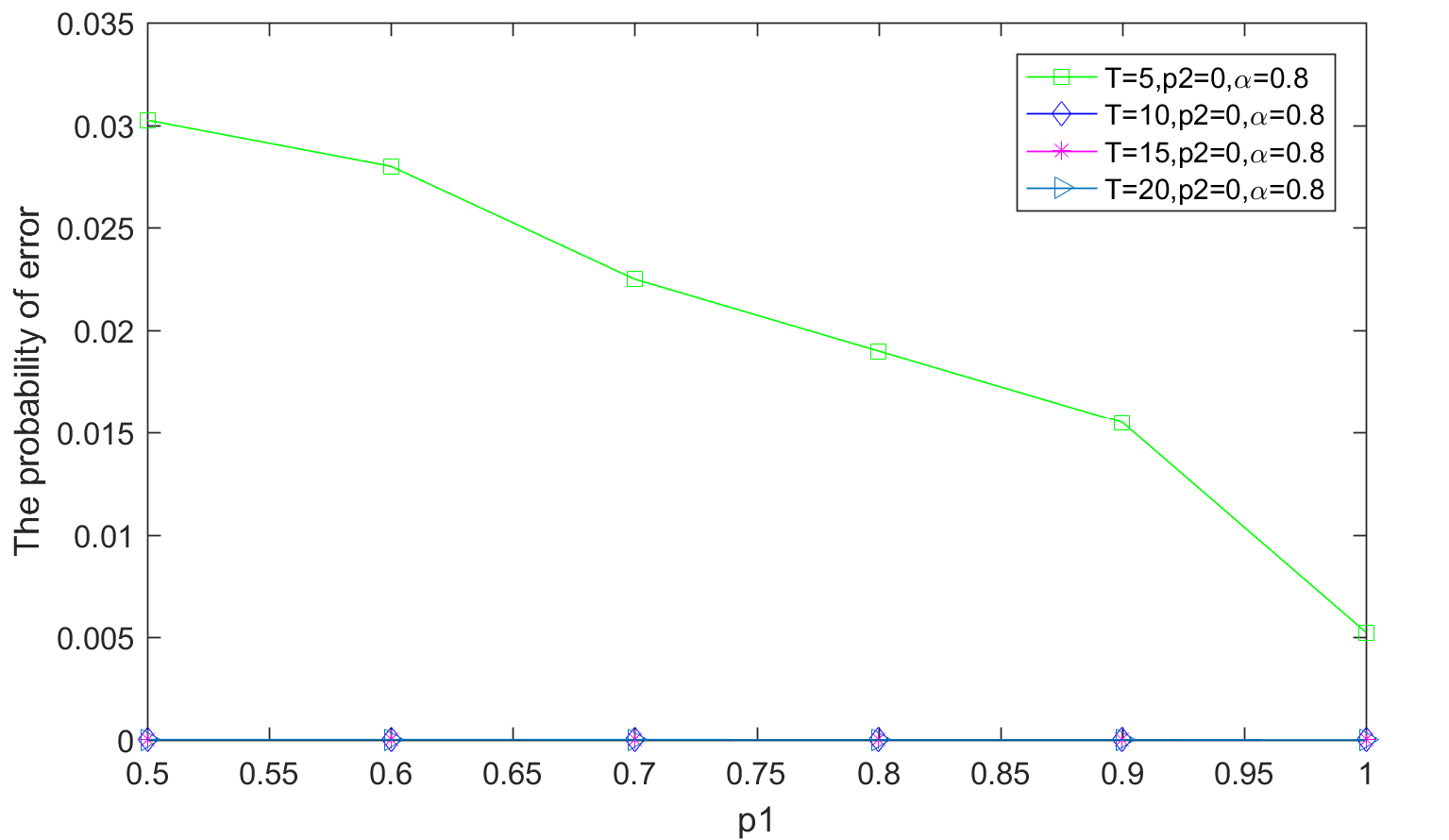}
  \caption{The probability of error versus the number of iterations given different value of $T$ for the proposed RACA with one anchor node when $\alpha_0=0.8$.}
  \label{Fig.main8}
\end{figure}

\begin{figure}[htbp]
\centerline{\includegraphics[width=17em,height=10em]{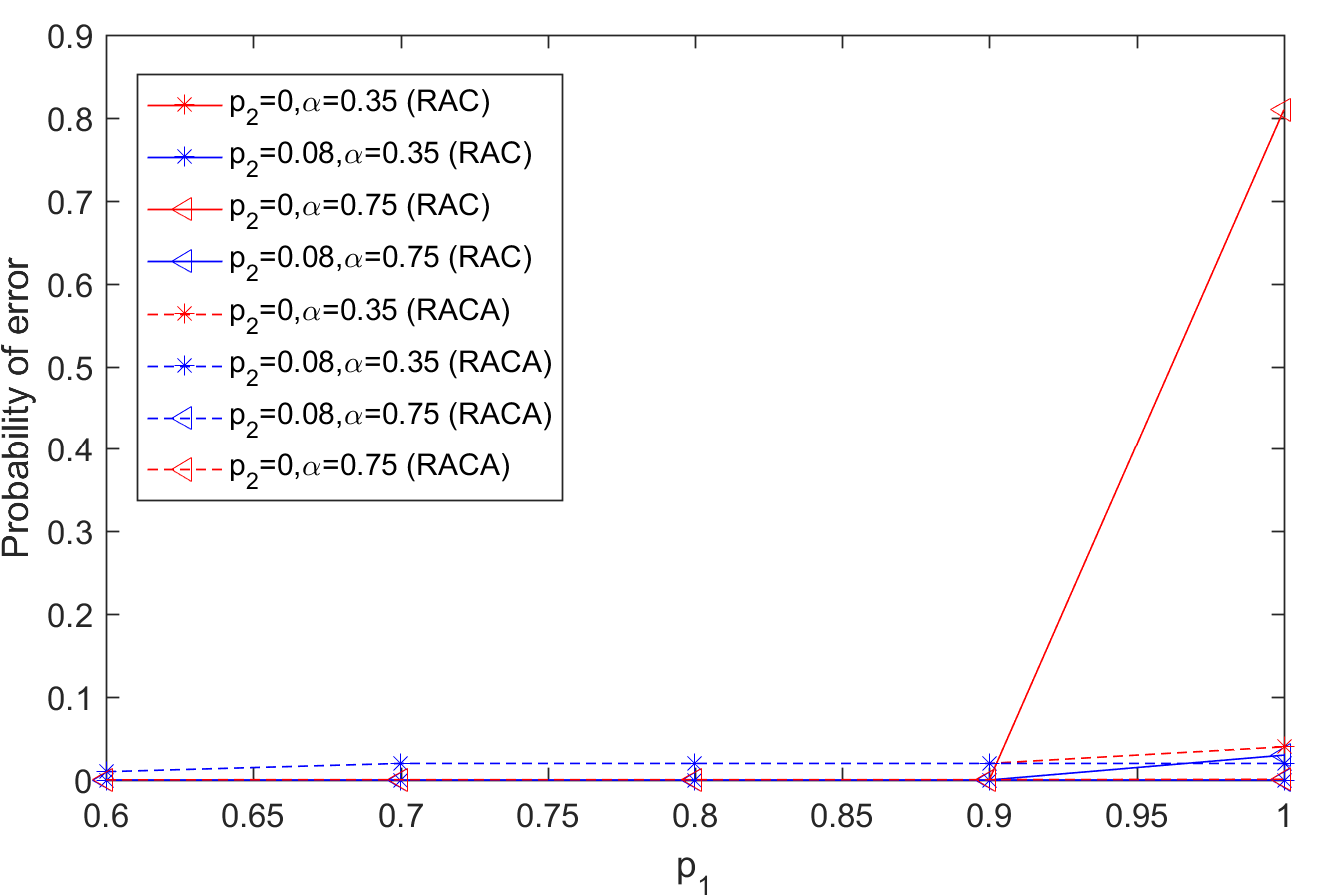}}
\caption{The probability of error versus dynamically changing $p_1$ given different value of $p_2$ and $\alpha_0$ for our proposed algorithms.}
\label{Fig.main7}
\end{figure}

In Fig. ~\ref{Fig.main5}(a), we can observe that the proposed algorithm with one anchor node (i.e., RACA) and the one without anchor nodes (i.e., RAC) both have outstanding detection performance when the fraction of Byzantine nodes $\alpha_0=0.35$. Fig. ~\ref{Fig.main5}(b) shows that the system can still obtain a good detection performance with the help of anchor nodes even when the fraction of Byzantine nodes is greater than $0.5$. It can be observed that the detection performance of the RAC degrades significantly only when $p_1$ approaches 1 and $p_2$ approaches 0 if the fraction of Byzantine nodes is greater than 0.5. We compare our proposed algorithms with other two algorithms in Fig. ~\ref{Fig.main5}(a). It shows that our proposed algorithms outperform Adaptive reputation clustering algorithm (ARC) proposed in \cite{vempaty2014false} and Reputation based algorithm (RBA) proposed in \cite{rawat2010collaborative}. Note that those algorithms (ARC and RBA) immediately break down when over half of the sensors are Byzantine nodes. In Fig.~\ref{Fig.main1}, we show that our proposed algorithms are able to quickly identify Byzantine nodes so that we can obtain an excellent detection performance. Furthermore, we examined the impact of the number of anchor nodes used on the performance of the system in Fig.~\ref{Fig.main2}. It shows that an increasing number of anchor nodes significantly enhances the detection performance of the system since a larger number of anchor nodes could provide the FC better reference decisions to identify the Byzantine nodes. In Fig. \ref{Fig.main8}, we show that at least 10 decision records, i.e., $T=10$, are needed to guarantee a relatively good detection performance. 

Fig.~\ref{Fig.main7} shows the detection performance of our proposed algorithms under Byzantine attacks with dynamically changing attacking parameter $p_1$. In each time step, we assume that the real value of $p_1$ is uniformly generated from $[p_1-0.05,p_1+0.05]$ in order to represent the dynamically changing attacking parameter $p_1$. The average error probability we obtain for a specific dynamically changing $p_1\in[p_1-0.05,p_1+0.05]$ is the error probability that corresponds to $p_1$ in Fig.~\ref{Fig.main7}.  We can observe that our proposed algorithms are able to defend against attackers whose attacking parameters are dynamically changing. It is due to the fact that the performance improvement of our proposed algorithms is directly affected by the deviation of Byzantine nodes' decision records from those of honest nodes, and the deviation is the reflection of $p_1$ and $p_2$ in the system. Hence, the dynamically changing attacking parameters in each iteration do not result in a significant impact on the ability of our proposed algorithms to defend against attacks.


\section{CONCLUSION}
In this paper, we proposed the RAC algorithm and the RACA algorithm to defend against Byzantine attacks in sensor networks when the FC is not aware of the attacking strategy. We utilized the history of local decisions and MMS results to update the reputation index of sensors and help the system accurately identify Byzantine nodes. Our simulation results showed that we are able to achieve superior detection performance and the enhanced ability of identifying Byzantine nodes by employing anchor nodes even when the Byzantines exceed half of the total number of sensors in the network. Furthermore, we showed that our algorithms are capable of defending against attackers whose attacking parameters change dynamically over time.
\appendices
\section{Proof of Lemma~\ref{lemma1}}\label{proof_lemma}
We have the following four cases when we consider the MMS results of the sensors in the same group\cite{}. Let $i$ and $j$ represent the sensors in the same group.
\paragraph{If $u_i=z_i$ and $u_j=z_j$} $i$ is a Byzantine node with probability

\begin{equation}\label{eq:alpha_LL}
\begin{split}
    \alpha_1=&P(i=B|u_i=z_i, u_j=z_j)\\
    =&\frac{\alpha_0^2f_{BB}^{(1)}+\alpha_0(1-\alpha_0)f_{BH}^{(1)}}{\alpha_0^2f_{BB}^{(1)}+\alpha_0(1-\alpha_0)(f_{HB}^{(1)}+f_{BH}^{(1)})+(1-\alpha_0)^2f_{HH}^{(1)}},
\end{split}
\end{equation}
where $f_{BB}^{(1)}=[2p_1p_2(1-p_1)+(1-2p_1+2p_1^2)(1-p_2)]^2$, $f_{HB}^{(1)}=f_{BH}^{(1)}=(1-p_2)(1-2p_1+2p_1^2)$ and $f_{HH}^{(1)}=1$.
\paragraph{If $u_i\neq z_i$ and $u_j=z_j$} $i$ is a Byzantine node with probability

\begin{equation}\label{eq:alpha_LU}
\begin{split}
    \alpha_2=&P(i=B|u_i\neq z_i,u_j=z_j)\\
    =&\frac{\alpha_0^2f_{BB}^{(2)}+\alpha_0(1-\alpha_0)f_{BH}^{(2)}}{\alpha_0^2f_{BB}^{(2)}+\alpha_0(1-\alpha_0)(f_{HB}^{(2)}+f_{BH}^{(2)})+(1-\alpha_0)^2f_{HH}^{(2)}},
\end{split}
\end{equation}
where $f_{BB}^{(2)}=[2p_1p_2(1-p_1)+(1-2p_1+2p_1^2)(1-p_2)][1-2p_1p_2(1-p_1)-(1-2p_1+2p_1^2)(1-p_2)]$, $f_{HB}^{(2)}=p_2(1-2p_1+2p_1^2)$,$f_{BH}^{(2)}=2p_1(1-p_2)(1-p_1)$ and $f_{HH}^{(2)}=0$.
\paragraph{If $u_i=z_i$ and $u_j\neq z_j$} $i$ is a Byzantine node with probability
\begin{equation}\label{eq:alpha_UL}
\begin{split}
    \alpha_3=&P(i=B|u_i=z_i,u_j\neq z_j)\\
    =&\frac{\alpha_0^2f_{BB}^{(3)}+\alpha_0(1-\alpha_0)f_{BH}^{(3)}}{\alpha_0^2f_{BB}^{(3)}+\alpha_0(1-\alpha_0)(f_{HB}^{(3)}+f_{BH}^{(3)})+(1-\alpha_0)^2f_{HH}^{(3)}},
\end{split}
\end{equation}
where $f_{BB}^{(3)}=[2p_1p_2(1-p_1)+(1-2p_1+2p_1^2)(1-p_2)][1-2p_1p_2(1-p_1)-(1-2p_1+2p_1^2)(1-p_2)]$, $f_{HB}^{(3)}=2p_1(1-p_2)(1-p_1)$,$f_{BH}^{(3)}=p_2(1-2p_1+2p_1^2)$ and $f_{HH}^{(3)}=0$.

\paragraph{If $u_i\neq z_i$ and $u_j\neq z_j$} $i$ is a Byzantine node with probability

\begin{equation}\label{eq:alpha_UU}
\begin{split}
    \alpha_4=&P(i=B|u_i\neq z_i,u_j\neq z_j)\\
    =&\frac{\alpha_0^2f_{BB}^{(4)}+\alpha_0(1-\alpha_0)f_{BH}^{(4)}}{\alpha_0^2f_{BB}^{(4)}+\alpha_0(1-\alpha_0)(f_{HB}^{(4)}+f_{BH}^{(4)})+(1-\alpha_0)^2f_{HH}^{(4)}},
\end{split}
\end{equation}
where $f_{BB}^{(4)}=[2p_1(1-p_2)(1-p_1)+p_2p_1^2]^2$, $f_{HB}^{(4)}=f_{BH}^{(4)}=2p_1p_2(1-p_1)$ and $f_{HH}^{(4)}=0$.

According to the above results, we have
\begin{subequations}
	\begin{align}
		\underline{\alpha}=P(i=B|i\in\underline{\mathcal{T}})&=\alpha_1\\
		\overline{\alpha}=P(i=B|i\in\overline{\mathcal{T}})&=\alpha_2P(u_i\neq z_i,u_j=z_j)\notag\\
		&+\alpha_3P(u_i=z_i,u_j\neq z_j)\notag\\
		&+\alpha_4P(u_i\neq z_i,u_j\neq z_j)
	\end{align}
\end{subequations}
The derivative of $\underline{\alpha}$ with respect to $p_1$ is given by (note that $f_{HB}^{(1)}=f_{BH}^{(1)}$)
\begin{equation}\label{eq:proof1}
		\frac{\partial\underline{\alpha}}{\partial p_1}=\frac{\frac{\partial\mathcal{F}_1}{\partial p_1}\mathcal{F}_2-\frac{\partial\mathcal{F}_1}{\partial p_1}\mathcal{F}_2}{\mathcal{F}_2^2}
\end{equation}
where $\mathcal{F}_1=\alpha_0^2f_{BB}^{(1)}+\alpha_0(1-\alpha_0)f_{BH}^{(1)}$, $\mathcal{F}_2=\alpha_0^2f_{BB}^{(1)}+\alpha_0(1-\alpha_0)(f_{HB}^{(1)}+f_{BH}^{(1)})+(1-\alpha_0)^2f_{HH}^{(1)}$, $\frac{\partial\mathcal{F}_1}{\partial p_1}=\alpha_0^2\frac{\partial f_{BB}^{(1)}}{\partial p_1}+\alpha_0(1-\alpha_0)\frac{\partial f_{BH}^{(1)}}{\partial p_1}$ and $\frac{\partial\mathcal{F}_2}{\partial p_1}=\alpha_0^2\frac{\partial f_{BB}^{(1)}}{\partial p_1}+2\alpha_0(1-\alpha_0)\frac{\partial f_{BH}^{(1)}}{\partial p_1}+(1-\alpha_0)^2\frac{\partial f_{HH}^{(1)}}{\partial p_1}$. Let $\frac{\partial\underline{\alpha}}{\partial p_1}=0$, we have
\begin{equation}\label{eq:p_1}
    \alpha_0^2\frac{\partial f_{BB}^{(1)}}{\partial p_1}f_{BH}^{(1)}+(1-\alpha_0)\frac{\partial f_{BB}^{(1)}}{\partial p_1}+(1-\alpha_0)^2\frac{\partial f_{BH}^{(1)}}{\partial p_1}=\alpha_0^2\frac{\partial f_{BH}^{(1)}}{\partial p_1}f_{BB}^{(1)}
\end{equation}
due to the fact that $\frac{\partial f_{HH}^{(1)}}{\partial p_1}=0$, where
\begin{align}
    \frac{\partial f_{BB}^{(1)}}{\partial p_1}=&8p_1(2p_2-1)^2(1-2p_1)(1-p_1)\notag\\
    &+4p_1(2p_2-1)(1-2p_1)(1-p_2)\\
    \frac{\partial f_{BH}^{(1)}}{\partial p_1}=&2(1-p_2)(2p_1-1).
\end{align}
We can easily obtain that $p_1=\frac{1}{2}$ makes the equation \eqref{eq:p_1} always hold for any specific $p_2\in[0,1]$. In other words, we can obtain that $p_1=\frac{1}{2}$ could minimize or maximize $\underline{\alpha}$ given a specify $p_2$. Correspondingly, $p_1=0\text{ or }1$ could also maximize or minimize $\underline{\alpha}$. Table \ref{tab:possible} shows all the possible maximum or minimum values of $\underline{\alpha}$ given a specific $p_2$.
\begin{table}[ht]
\caption{Possible maximum or minimum values of $\underline{\alpha}$ given a specific $p_2$.}
\centering
\begin{tabular}{|c|c|c|c|}
\hline
        & $f_{BB}^{(1)}$ & $f_{BH}^{(1)}$ & $f_{HH}^{(1)}$ \\ \hline
$p_1=0$   & $(1-p_2)^2$    & $1-p_2$        & $1$            \\ \hline
$p_1=1/2$ & $1/4$          & $(1-p_2)/2$    & $1$            \\ \hline
$p_1=1$   & $(1-p_2)^2$    & $1-p_2$        & $1$            \\ \hline
\end{tabular}
\label{tab:possible}
\end{table}
According to Table \ref{tab:possible}, we have
\begin{equation}\label{eq:ref_under}
    \underline{\alpha}=\frac{\alpha_0^2/4+\alpha_0(1-\alpha_0)(1-p_2)/2}{\alpha_0^2/4+\alpha_0(1-\alpha_0)(1-p_2)+(1-\alpha_0)^2}
\end{equation}
for $p_1=\frac{1}{2}$ and
\begin{equation}\label{eq:ref_over}
    \underline{\alpha}=\frac{\alpha_0^2(1-p_2)^2+\alpha_0(1-\alpha_0)(1-p)}{\alpha_0^2(1-p_2)^2+2\alpha_0(1-\alpha_0)(1-p_2)+(1-\alpha_0)^2}
\end{equation}
for $p_1=0$ or $p_1=1$. Let $h=\underline{\alpha}-\alpha_0$ represent the difference between $\underline{\alpha}$ and $\alpha_0$, and it is given by
\begin{equation}
    h=\frac{\alpha_0(1-\alpha_0)(5\alpha_0-4+2(1-2\alpha_0)(1-p_2))}{4(\alpha_0^2(1-p_2)^2+2\alpha_0(1-\alpha_0)(1-p_2)+(1-\alpha_0)^2)}
\end{equation}
for $p_1=\frac{1}{2}$, and it is given by
\begin{equation}
    h=\frac{\alpha_0(1-\alpha_0)(\alpha_0-1+(1-2\alpha_0)(1-p_2))}{\alpha_0^2/4+\alpha_0(1-\alpha_0)(1-p_2)+(1-\alpha_0)^2}
\end{equation}
for $p_1=0$ or $p_1=1$. Because we only care about the sign of $h$ and the numerator is always positive. Let $h_d$ denote the  numerator of $h$. We have
\begin{equation}\label{eq:pf_derivertive}
    \frac{\partial h_d}{\partial p_2}=\left\{
    \begin{array}{rcl}
      \frac{\alpha_0(1-\alpha_0)(2\alpha_0-1)}{2}&\text{,if $p_1= \frac{1}{2}$}\\
      -\alpha_0(1-\alpha_0)(1-2\alpha_0)&\text{,if $p_1= 0$ or $p_1= 1$}
    \end{array}\right.
\end{equation}
We can easily obtained that $\frac{\partial h_d}{\partial p_2}\leq0$ for $\alpha_0\leq\frac{1}{2}$ and $\frac{\partial h_d}{\partial p_2}\geq 0$ for $\alpha_0\geq\frac{1}{2}$ given $\forall p_1\in\{0,\frac{1}{2},1\}$. According to \eqref{eq:pf_derivertive}, We can prove that $h_d\leq0$ always holds for $\forall p_1\in\{0,1\}$. When $p_1=\frac{1}{2}$, $h\leq0$ also holds for $\forall \alpha_0\in[0,\frac{1}{2}]$. 

\begin{figure}
    \centering
    \includegraphics[width=17em,height=10em]{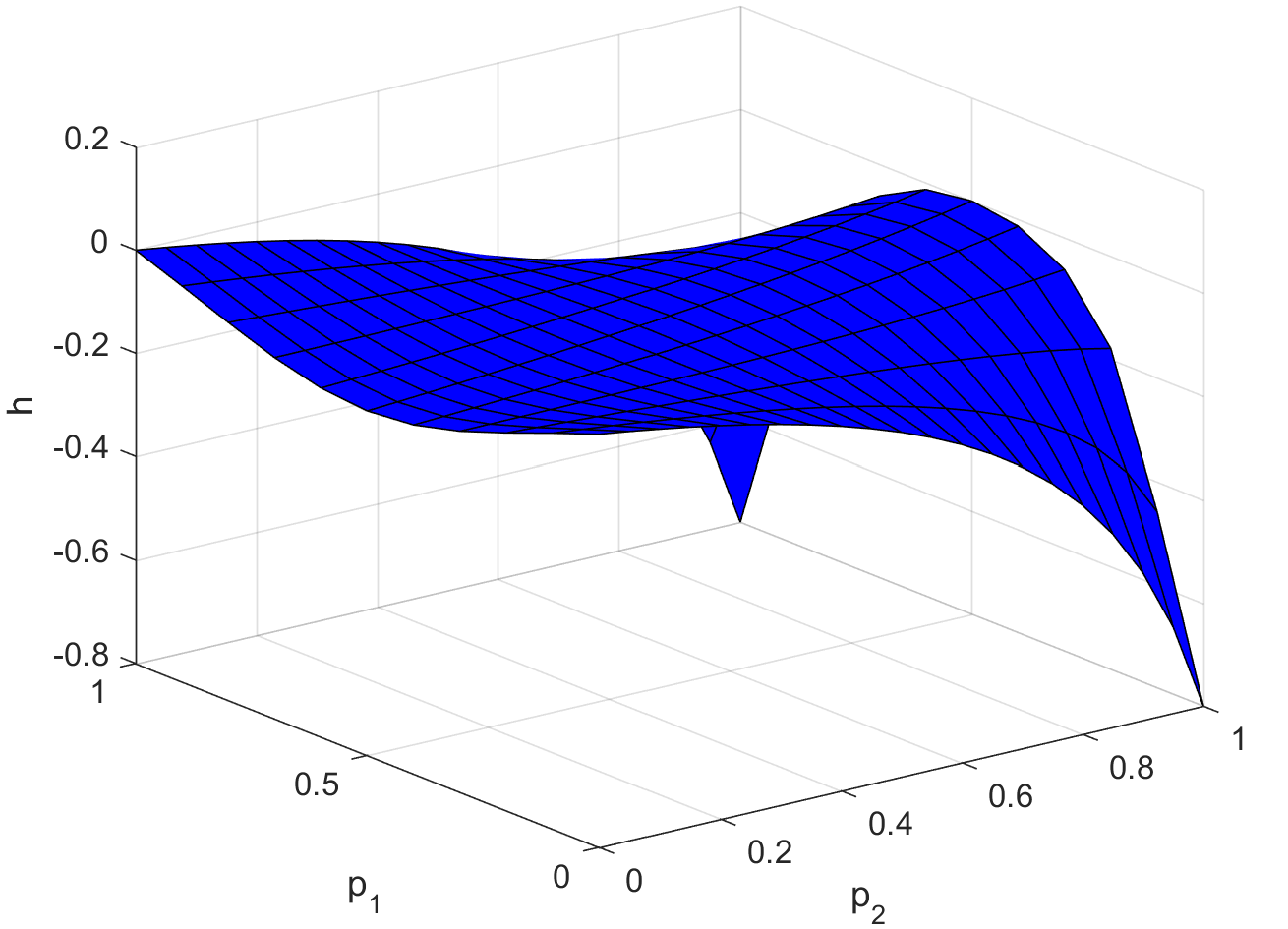}
    \caption{$h$ versus $p_1$ and $p_2$ given $\alpha_0=0.8$.}
    \label{Fig.main6}
\end{figure}

When $\alpha_0>\frac{1}{2}$ and $p_1=\frac{1}{2}$, the value of $p_2$ that guarantees $h\leq0$ should be smaller than $p_2^{max}$. $p_2^{max}$ can be obtained from letting $\frac{\partial h}{\partial p_2}=0$ and it is given by
\begin{equation}\label{eq:max_p}
    p_2^{max}=\min\left\{\frac{\alpha_0-2}{2(1-2\alpha_0)},1\right\}
\end{equation}
for $\alpha_0>\frac{1}{2}$ and $p_1=\frac{1}{2}$. Note that  $p_2^{max}=1$ implies $\frac{\alpha_0-2}{2(1-2\alpha_0)}\ge 1$ such that $\alpha_0\le 0.8$. Hence, if the fraction of Byzantine nodes in the network is smaller than 0.8, i.e., $\alpha_0\leq0.8$, $\underline{\alpha}$ is always smaller than $\alpha_0$. In other words, we always have a lower probability of existence of Byzantine nodes in set $\underline{\mathcal{T}}$ when $\alpha_0\leq0.8$. The detailed proof for this part is omitted here. Fig. \ref{Fig.main6} corroborates the results in \eqref{eq:max_p}.

Since $P(i\in\underline{\mathcal{T}})\underline{\alpha}+P(i\in\overline{\mathcal{T}})\overline{\alpha}=\alpha_0$, we have
\begin{equation}\label{eq:range}
	\begin{split}
	P(i\in\underline{\mathcal{T}})\alpha_0+P(i\in\overline{\mathcal{T}})\overline{\alpha}&\geq \alpha_0\\
	P(i\in\overline{\mathcal{T}})\overline{\alpha}&\ge\alpha_0(1-P(i\in\underline{\mathcal{T}})\\
	\overline{\alpha}&\geq \alpha_0
	\end{split}
\end{equation}
when $p_2\in[0,p_2^{max}]$, $\alpha_0\in[\frac{1}{2},1]$ and $p_1\in[0,1]$. According to the above analysis, \eqref{eq:range} also holds when $\alpha_0\in[0,\frac{1}{2}]$, $p_1\in[0,1]$ and $p_2\in[0,1]$.
\bibliography{refer.bib}
\bibliographystyle{IEEEtran}
\end{document}